\documentclass[11pt]{article}

\usepackage{float}
\usepackage[a4paper, total={6in, 8in}]{geometry}
\usepackage{amssymb, amsmath, amsthm}
\usepackage{threeparttable}
\usepackage{booktabs}
\usepackage{setspace}
\usepackage{graphicx}
\usepackage{natbib}
\usepackage[colorlinks=true,linkcolor=black,anchorcolor=black,citecolor=black,filecolor=black,menucolor=black,runcolor=black,urlcolor=black]{hyperref}

\begin{document}

\onehalfspacing
\title{\Large \bf Estimating granular house price distributions in the Australian market using Gaussian mixtures}

\author{ 
\begin{tabular}{ccccccc}
\textbf{\small Willem P Sijp} &  & \textbf{\small Anastasios Panagiotelis}  \\%
[0.1in] 
{\footnotesize Neoval Pty Ltd \&} &  & {\footnotesize University of Sydney }  \\
{\footnotesize University of Sydney} &  &  
\end{tabular}
}

\maketitle
\thispagestyle{empty}

\bigskip

\begin{abstract} 
A new methodology is proposed to approximate the time-dependent house price distribution at a fine regional scale using Gaussian mixtures. The means, variances and weights of the mixture components are related to time, location and dwelling type through a non linear function trained by a deep functional approximator. Price indices are derived as means, medians, quantiles or other functions of the estimated distributions. Price densities for larger regions, such as a city, are calculated via a weighted sum of the component density functions. The method is applied to a data set covering all of Australia at a fine spatial and temporal resolution. In addition to enabling a detailed exploration of the data, the proposed index yields lower prediction errors in the practical task of individual dwelling price projection from previous sales values within the three major Australian cities.  The estimated quantiles are also found to be well calibrated empirically, capturing the complexity of house price distributions.
\end{abstract}



\bigskip

\clearpage

\doublespacing

\section{Introduction}\label{sec1}

House price indices are an important tool in government economic decision making and can assist in the valuation of the collateral behind mortgage portfolios by market agents such as brokers.
The hedonic and the repeat sales approach are the most entrenched house price index methodologies in practical use today. They track the change in mean price over time, typically at the metropolitan scale. This can be misleading when there is significant variation in appreciation within a metropolitan region. Indeed, variation in house price appreciation within cities is large, as shown by e.g. \citet{guerrieri_et_al2013, bogin_et_al2019,landvoigt_et_al2015}, indicating that fine-scale indices are more useful \citep[e.g.][]{ constantinescu_francke2013, bogin_et_al2019, ren_et_al2017, waltl2019}.  For Sydney, \citet{waltl2019} show that the rise and subsequent decline of the 2004 boom were significantly more pronounced for Sydney's low-priced outer suburbs than the high-priced areas closer to the city.  Therefore, a metro city-level index may be too coarse to estimate current loan-to-value ratios for mortgages \citep{bogin_et_al2019}. These concerns have helped stimulate the development of geographically fine-scaled indices. This is challenging, as real estate tends to be simultaneously very heterogeneous and infrequently traded \citep{deng_et_al2012}.

Several approaches have been proposed to address the data sparsity issues that arise from infrequent trading of property at a fine spatial scale. \cite{bogin_et_al2019} use an annual repeat sales index and formulate indices on the American ZIP code level. That time step is relatively coarse, illustrating a fundamental tradeoff between temporal- and spatial resolution due to increased noise at the granular level \citep{geltner_ling2006}. 
 As such, using the whole dataset to estimate spatially localised trends is helpful, as shown by \citet{francke_devos2000} and \cite{francke_minne2017}. 
 Similarly, \cite{ren_et_al2017} leverage cluster similarities based on neighbourhood price dynamics to yield cluster indices for Seattle. These are fine geographical scales, where data naturally becomes sparse when a specific area is considered in isolation, and these methods address the issue of spatial and temporal data sparsity.

Another limitation of the original hedonic and repeat sales methods is that they do not yield information about the underlying cross-sectional price distributions that gave rise to the index time series. The usefulness of distributions in gaining a richer picture of the market has led to new work to fill this lacuna and the study of distributions and their quantiles is now an active area of new research \citep{waltl2019, nicodemo_raya2012}. Recent studies, including \citet{coulson_mcmillen2007, deng_et_al2012, mcmillen2008, mcmillen_shimizu2021, waltl2019,nicodemo_raya2012} employ quantile regression and/ or kernel density approaches to estimate a cross-sectional house price distribution. See \cite{waltl2019} for a review of studies that use quantile regression to elucidate cross sectional distributions, including approaches that allow regression coefficients to vary over the distribution and space. 

Here, we propose to target the price distribution of data clusters employing an analytical function. We perform a parametric Gaussian mixture estimation of the conditional cross-sectional probability density distributions of price at each point in time for each property type and each (statistical area level 3, SA3, see below) subregion.  Distributions more generally, such as distributions over metropolitan areas, are obtained by weighted sums of the granular fitted distributions using time-invariant weights representing a fixed population of houses. The Gaussian mixtures yield flexible probability density functions and may be able to accomodate the effect of subpopulations associated with unresolved model features and spatial price variation within a subarea. The fitting task is performed by a deep neural network, but the approach is intended to be independent of the details of the function approximator for the density function parameters.

We now discuss some closely related work. \cite{waltl2019} construct location- and segment-specific house price indices using a quantile imputation approach. Their hedonic model includes a spline term dependent on latitude and longitude, allowing quarterly quantile time series to be constructed depending on space (with fixed regression coefficients).
They construct cross-sectional price densities and uncover spatial variation in appreciation patterns.
\citet{mcmillen_shimizu2021} examine the effect of compositional changes using a quantile regression for attribution (also including spatial features) alongside a second model comprising a mixture of annual kernel density estimates weighted by a logit sale time propensity model to emulate the housing mixture of the initial reference period to fix characteristics \citep{hill_melser2008} in their yearly time step index. In contrast to these approaches, we use a parametric distribution, the Gaussian mixture, to approximate the cross section price distribution.

Recent studies have examined the use of Gaussian mixtures in determining submarkets in specific regions.
One of the perhaps more foundational papers on market mixture models is \cite{goodman_thibodeau_jhe1998}, who examine submarkets in terms of a mixture of hedonic functions and a review of this topic is found in \cite{islam_rurds2009}.
Among significant examples of further developments are the models of  \cite{belasco_jrer2012} and \cite{nishi_jhe2021}. In brief, \cite{belasco_jrer2012} use an expectation maximisation algorithm to determine optimal parameters that themselves are hedonic linear functions. Their model expresses the various preferences differentiated among the individual buyers, who are found to fall into three groups based on demography in their study. In contrast, \cite{nishi_jhe2021} use an automated segmentation approach that actually also predicts the conditional distribution of price and house characteristics. This approach obviates the need for experiments to determine the number of mixture components, instead using the stick breaking algorithm. These studies are similar to ours in that they employ density mixtures. However, they differ from ours in both focus and the employed methodology.  \cite{belasco_jrer2012} identify submarkets within a small confined geography (820 dwellings) and do not consider time as a variable. Instead, their focus is the identification of submarkets within the region and interpreting this in terms of external data, meaning that the hedonic coefficients are important in their study. Namely, their mixture densities are conditional on hedonic linear functions so as to be able to determine the value that each submarket places on a hedonic good, such as square footage of the house: these coefficients can take multiple values, as a mixture would allow.  This focus differs from our study, where time is the primary focus: we are interested in constructing time-indexes.
Also,  our data volumes are relatively large and we examine regions that differ strongly in their time-behaviour. For instance, the decadal time scale of our data means that doubling or tripling of house values is the norm, and even within a single city, spatially the price variation may approach an order of magnitude. The variations in this are part of our focus.
Also, unlike the mixture density network in our study, the finite mixture model of \cite{belasco_jrer2012} employs an expectation maximisation algorithm first developed in \cite{dempster_laird_rubin1977} as applied to Gaussian mixtures: this approach is further described in Chapter 9.3 of \cite{bishop2006}.

Our paper also adds to the growing literature on using machine learning in property applications. For example \citet{kok_et_al2017} use non-linear approaches to obtain individual house price predictions, and Automated Valuation Models (AVM) based on machine learning are now routinely used commercially. If spatially detailed inputs are used, AVM outputs could be averaged to yield detailed local price indices and even distributions. Principled aggregation methods exist: the model-agnostic framework of \citet{calainho_et_al2022} allows individual predictions from machine learning models (among others) to be leveraged to produce a price index using a single imputation Chained Paasche approach. They test their framework for XGBT, SVR, avNet and Cubist. In contrast to using detailed AVM output to construct a mean index or estimate sample price distributions, to the best of our knowledge, our proposed approach is the first to use machine learning to estimate house price distributions parametrically, rather than simply the mean or sample statistics based on AVM output. 

In a sense, our paper is about unobserved heterogeneity in housing market. Our focus is not on the details of this. Instead, we view the densities discussed in this paper as general distributions without further interpretation.  For a deeper investigation into this topic of unobserved heterogeneity, we refer to \cite{francke_ree2021} who construct a per-house random effect model within a hedonic estimation that yields significant improvements in individual house price valuation. Their study has a focus on random effects and individual price valuation and not on price index behaviour in time, and therefore has a different angle compared to our work.

In terms of methodologies to obtain house price distributions over time, we point to the well-known work of \citet{mcmillen2012}. Here a matching-based method is used to track cross-section price density and a principled way to increase sample size in generalised repeat sales methods. They match properties that satisfy a reasonable matching criterion, in their case using a logit propensity model for the time of sale. In each quarter retaining a constant-sized population that is individually matched to the chosen base period population, they track any quantile of the constant size population in each quarter. Similar to the hedonic method, this approach cannot match on unobservables.  
While aiming to achieve similar ends, our approach differs significantly from this work, primarily in the way that mixture distributions and machine learning methods are used to derive indices at a much finer spatial and temporal scale compared to existing work. 

For our empirical work, the entire dataset for Australia is used in each single model run to determine the granular densities, allowing for the implicit identification of major price trends and perturbations or variations thereon within the model. Mean price indices are derived from the Gaussian mixture densities. The model yields low-noise time series of average price recorded weekly. A benefit of this temporal granularity is that it allows gradients to be more accurately presented in the model. The time-input smoothing smears values across weeks. 
As might be expected, we find that indices defined on smaller regions yield lower prediction errors in the practical task of individual dwelling price projection from previous sales value for the three major Australian cities. This adds to the findings in the literature that the geographical variation in price appreciation matters in indexation.

To illustrate in some detail, applying the index to small geographical regions shows that price appreciation has lagged behind Sydney in two desirable but affordable regions further away from Sydney, the Blue Mountains and the beach destination of Coffs Harbour, from 2012 up to 2017 and have caught up since then during a phase of accelerated growth. Major market booms and declines in the last 4 years began concurrently for these regions. Also, the more expensive percentiles have seen more volatile growth and decline than the cheapest percentiles in the Sydney metro region.

The rest of the paper is ordered as follows. Section~\ref{formulation} introduces the details of the new model (Section~\ref{sec:dmodel}) and a hedonic and repeat sales model for comparison  (Section~\ref{sec:linear}). Section~\ref{sec:results} discusses the results of the model runs we have conducted in terms of its outputs. This comprises examples of probability densities at the subregional and city level (Section~\ref{sec:result_pdf}), 
the time series indices derived from these probability densities (Section~\ref{sec:result_series}). In Section~\ref{sec:performance} we validate the model in terms of individual dwelling appreciation prediction and median estimation. 
Finally, Section~\ref{sec:discussion} contains a discussion of the results and their implications.

\section{Price Index Construction}
\label{formulation}

Here we describe the data, the model and the two linear benchmark indices based on the hedonic and the repeat sales regressions. Boldface variables indicate vectors and matrices, whereas lightface variables are scalars.

\subsection{Data}
\label{sec:subregions}

The data we use are Australian house sales sourced from Valocity Pty Ltd with dwelling prices measured in Australian dollars. 
To illustrate the robustness of our proposed methodology, the training data for the model have not been filtered by outlier detection, and so we tolerate some erroneous input entries. (Note that the linear indices introduced as benchmarks in Section~\ref{sec:linear} are fitted on cleaned data.) A distinction is also made between property types of ``house'' and ``unit'', which is handled via the inclusion of a dummy variable in the model. The type “unit” refers to an apartment or a “townhouse” (as defined under the Australian building code). It can be thought of as a property within a complex of three or more dwellings with shared ownership of the land and common property. For brevity, we will omit mention of this input (property type) unless needed. 

Data are measured at a weekly frequency from 1990-01-01 to 2022-05-01 with data loaded from the database at the end date. In Section~\ref{sec:result_series} we will use updated data with the more recent end point of 2023-15-01 due to interesting market developments that became apparent since the inception of this paper earlier in 2022. For the spatial component, we use the Australian Statistical Areas Level 3 (SA3) region defined in the Australian Statistical Geography Standard (ASGS) used by the Australian Bureau of Statistics (ABS) and described in \cite{abs:2021}. The SA3 type of region represent areas serviced by major transport and commercial hubs when located in a major city. They generally contain a population of between 30,000 and 130,000 people and roughly 5 to 10 postcodes. There are 358 SA3 regions covering all of Australia without gaps or overlaps, we will refer to these as ``subregions'' in the following, omitting the ``SA3''. The subregions are concentrated in the three largest cities, with 39 in Brisbane, 40 in Melbourne and 47 in Sydney.
 Also, our approach works well at a finer geographical regional resolution, such as postcode and SA2 (not shown)\footnote{The SA2 have a population range of 3,000 to 25,000 persons and an average population of about 10,000 persons}.

\begin{table}[h!]
	\centering
	
\begin{tabular}{lrrr}
\toprule
     City &  Monthly Region &  Monthly City &  All (m) \\
\midrule
   Sydney &             142 &          6519 &      2.5 \\
Melbourne &             150 &          6013 &      2.3 \\
 Brisbane &              88 &          3403 &      1.3 \\
 Adelaide &              81 &          1506 &      0.5 \\
    Perth &             162 &          3405 &      1.3 \\
   Hobart &              47 &           280 &      0.1 \\
   Darwin &              34 &           136 &      0.1 \\
      ACT &              72 &           558 &      0.2 \\
\bottomrule
\end{tabular}

	\caption{Sales data broken down by major city for the model training period of 1990-2022. Average monthly sales by SA3 subregions within each city (column 2, ``Monthly Region''), average monthly sales for each state capital city (column 3, ``Monthly City'') and total sales of each city over the entire period (column 4, ``All (m)'', in million dollars). The model is trained on 12.3 million sales, of which 8.4 million occur in these capital cities.}
	\label{table:counts}
\end{table}

The model is fit on 12.34 million individual sales transaction records comprising a national sales database for the period 1990-2022, of which 8.4 million occur in the state capitals. Table~\ref{table:counts} shows a breakdown of these sales by capital city, including average monthly sales per SA3 subregion of that city, total monthly sales of that city and all sales for the entire period. In the three largest cities, monthly sales in an SA3 region average 88 for Brisbane, 142 for Sydney, and 150 for Melbourne. Average monthly sales are 3403 for Brisbane, 6013 for Melbourne and 6519 for Sydney.

\subsection{Density estimation and index}
\label{sec:dmodel}

We model the log price density in a subregion and construct the price index over a larger area as a weighted sum of its component subregions. 

\subsubsection{Log price density in a subregion}

We model the probability density function (PDF) $f(y \vert \mathbf{x}, T = t)$ of log sale prices $y$ for a subregion $r$ as a mixture of Gaussian distributions:

\begin{equation} \label{eq:mixture}
	f(y \vert \mathbf{x}, T = t) = \sum_{k=1}^{K} \alpha_k (\mathbf{x}, t;\mathbf{w}) \phi(y;\mu_k(\mathbf{x}, t;\mathbf{w}),\sigma^2_k(\mathbf{x}, t;\mathbf{w}))  \,,
\end{equation}

where $\mathbf{x} = (r, \mathbf{z}')'$ is the vector of a location indicator $r$ and discrete house characteristics $\mathbf{z}$ and $t$ is a time period. In this paper, $\mathbf{z}$ only includes the property type (unit or house) except in Section~\ref{sec:compositional}, where we examine the effect of including bedrooms and discretised log land area. 
The distribution parameters are estimated via the outputs of a three layer neural network. This type of mixture density network that we use is described in Chapter 5.6 of \cite{bishop2006}.
 The terms $\mu_k (\mathbf{x}, t;\mathbf{w})$, $\sigma^2_k (\mathbf{x}, t;\mathbf{w})$ and $\alpha_k (\mathbf{x}, t;\mathbf{w})$ are respectively the mean, variance and component weight of the $k^{th}$ mixture component and $\phi(y;\mu,\sigma^2)$ is the density of a Gaussian distribution with mean $\mu$ and variance $\sigma^2$. Naturally, the component weights are such that each density integrates to one, enforced by a softmax parameterisation. 
 The mean of each component  $\mu_k (\mathbf{x}, t;\mathbf{w})$ is the output of the function approximator that depends on model parameters $\mathbf{w}$. The functions for the variances and component weights, are similarly defined. Dropping the model parameters in the notation on the rhs for simplicity, the likelihood of the model is given by

\begin{equation} \label{eq:lik}
	L(\mathbf{w};\mathbf{y},\mathbf{x}, t)=\prod_{t=1}^T\prod_{\mathbf{x} \in \mathcal{I}_{\mathbf{x}}}\prod_{i} D_{i, \mathbf{x}, t} \sum_{k=1}^K\alpha_k(\mathbf{x}, t)\phi(y_i;\mu_k(\mathbf{x}, t),\sigma_k^2(\mathbf{x}, t))
\end{equation}

where $D_{i, \mathbf{x}, t}$ is an indicator function taking the value $1$ when sale $i$ has features $\mathbf{x}$ and took place at time $t$, and zero otherwise. As is common, maximum likelihood estimation is implemented by minimising the negative log likelihood loss, $-log L(\mathbf{w};\mathbf{y},\mathbf{x}, t)$. 
As price variations arise to a large extent (although not exclusively) from property type and location, we resolve SA3 subregion as a categorical variable in Eq.~\ref{eq:lik}.

 Time inputs are discretised as a categorical variable representing the number of weeks elapsed since 1 Jan 1990. The cyclical week within the year is added as an input to allow the function approximator to correct for a slight seasonality that is otherwise present in the index around the Christmas and new year period (arising from much more pronounced seasonality at Christmas in the data). Categorical inputs such as time are represented as dense vectors (with trainable coordinates) in a 10 dimensional vector space over the real numbers, following the common of approach of an embedding layer. As such, time input is an integer that is then converted to a time dummy that is then mapped to a dense vector (in practice via a lookup).  Gaussian noise with a standard deviation of two weeks, and followed by rounding to integer, is applied to this categorical time input variable during model runtime, enhancing covariance between adjacent weeks. This introduces a fuzzy time-window with an effective width of about a month, enhancing the effective number of data points seen by the model so as to encode proximity in time. Categorical region variables enter the model simultaneously with their neighbours. In this way subregions can exert an influence on the model responses to their neighbouring regions. As a result, regions with low levels of data benefit from information about neighbouring regions, as well as the more general main trends. No other smoothings are applied to the input and output values. 

Our model employs eight mixture components, $K=8$. In fact, checks using the AIC criterion show that AIC decreases significantly with increasing $K$, but this decline flattens out at $K=3$  with only very small improvements after that (not shown), suggesting diminishing returns for higher values.
 However, close examination of the distributions suggest that in some cases small regional details that are of interest may be captured at higher $K$ and we feel this higher value is sufficiently expressive to model a wide range of distributions, while still avoiding more complexity than needed (AIC does not increase here). The Gaussian mixture accommodates a range of probability distributions that could differ significantly from a single Gaussian, thus removing the misspecification bias associated with the single fixed Gaussian that is (in effect) used in common regression. There is also a possibility that the mixture reflects unresolved house characteristics.

Ensemble model averages of around 30 models, each initiated with different values of $\mathbf{w}$ are used to construct the indices by normalised pooling of the output Gaussians (resulting in another Gaussian mixture distribution with more components). In addition to regularisation methods applied to the fitting network the use of ensemble-averaging acts to regularise further by removing higher frequency serially uncorrelated random effects and the more temporary compositional biases associated with week to week sampling. This yields stabilised low-noise indices, as seen in Section~\ref{sec:results} below, based on distributions that reflect the underlying data (Section~\ref{sec:median}). Overall, the results are generally robust to slightly different choices of model architecture.  

The model is fitted on the entire dataset for Australia from January 1990 to May 2022 to obtain the time-evolution of the cross-section price density for each subregion.  Estimating individual component cross-sectional distributions and their time evolution, allows the implicit identification of main trends and perturbations thereon, an advantage emphasised and addressed in \citet{francke_minne2017, ren_et_al2017}. Once cross-sectional densities are obtained, central tendency- and segment specific measures such as the mean and median, as well as quantiles more generally, are obtained from these Gaussian mixture density functions, yielding time series, e.g. the average log price $\bar{y}_t = \mathbb{E}f(y \vert T = t)$.

\subsubsection{The counterfactual metropolitan index as a weighted sum}

We consider the log price density $f(y \vert T = t)$ of a metropolitan area as composed as a weighted sum of the densities of the subregions $r$ that make up the metropolitan area as:

\begin{equation} \label{eq:general_decomp}
	f(y \vert T = t) = \sum_{\mathbf{x} \in \mathcal{I}_\mathbf{x}} f(y \vert \mathbf{x}, T = t) h(\mathbf{x} \vert T = t)d\mathbf{x}	
\end{equation}

 where the weight $h(\mathbf{x} \vert T = t)$ is the probability of observing $\mathbf{x}$ in the sales at time $t$ and $\mathcal{I}_\mathbf{x}$ the cartesian product of the collections of distinct values the individual features $x_i$ can take. 
In integral from (denoting aggregation of both discrete and continuous inputs), \citet{mcmillen_shimizu2021} use the general decomposition shown in Equation \ref{eq:general_decomp}, including other features such as bedrooms etc instead of $r$, to arrive at ``counter-factual'' distributions by replacing the time-dependent weights $h$ with fixed estimates from their logit time-of-sale propensity model and using yearly general kernel density estimates for the conditional price distributions. This weight replacement reduces the effect of compositional changes of observables over time.

We replace the $h(\mathbf{x} \vert T = t)$ with time-invariant weights to track a fixed population, similar in general intention to \citet{mcmillen_shimizu2021}. We will simply use $h(\mathbf{x})$ for the fixed time period between Jan 1990 and Jan 2020 as weights so that the tracked population may reflect the total housing stock observed over that period. Namely, $h(\mathbf{x})$ is the probability density of observing the feature combination $\mathbf{x}$ in the population over the entire index period, estimated as: 

\begin{equation} \label{eq:weights}
	h(\mathbf{x}) = \frac{n_\mathbf{x}}{ \sum_{\mathbf{x} \in \mathcal{I}_\mathbf{x} } n_\mathbf{x} }
\end{equation}

where $n_\mathbf{x}$ denotes the number of observations with characteristics and location categories $\mathbf{x}$. As a justification for this, we intend the weights to reflect the total housing stock as it has been observed towards the end of the period. Namely, being based on most of the observed sales, it is an approximation of this quantity.
  This makes the index more useful for more recent (and therefore more relevant) time intervals.
We simplified our model by only considering a limited discrete collection for $\mathcal{I}_\mathbf{x}$, so that many houses share the same $\mathbf{x}$, tracking only a relatively small number of component distributions $f(y \vert \mathbf{x}, T = t)$.


\subsection{Benchmarks}
\label{sec:linear}

For comparison to other methodologies, we compare our indices with a hedonic model and a repeat sales method. We describe each of these in turn. We will refer to our model and derived indices as D-model and D-index. For succinctness, we will refer to the cross-section (log) price distributions as densities. A separate density is available at each \emph{specific time}. We will use three derived indices from our D model: the ``D-median'' index, the median of the composite metro PDF, ``D-gmean'' index, the geometric mean of that metro PDF and the ``D-subregion''  index, the median of the regional level PDF.
All indices we consider and their features are summarised in Table~\ref{table:indices}.

\begin{table}[h!]
	\centering
	\begin{tabular}{llll}
		\toprule
		Index Name &  Description & Composite & Domain  \\
		\midrule
		Hedonic &  hedonic index & no & metro     \\
		Repeat Sales &  repeat sales & no & metro       \\

		D-median &  median of PDF & yes & metro      \\
		D-gmean &   geom mean of PDF & yes & metro   \\

		D-subregion &  Median of PDF & no & subregion      \\

		\bottomrule
	\end{tabular}
	\caption{Summary of value-based indices discussed in this paper. The ``domain'' column indicates the domain on which the index is defined, where ``metro'' indicates metropolitan area, e.g. Sydney. The ``composite'' column indicates whether the index has been composed from subregion PDFs. The group of D metro indices derives from one single underlying metropolitan PDF model, whereas the finer scale D-subregion index derives from the PDF for the individual subregions. }
	\label{table:indices}
\end{table}

\subsubsection{Hedonic Index}

The hedonic methodology \citep{rosen:1974} explicitly resolves house characteristics and locational features. A downside is that fewer features may be available than one desires and this could create a biased estimate \citep[e.g.][]{ekeland_et_al2004}. Hedonic imputation approaches, e.g. Laspeyres and Paasche indices, are thought to difference out such bias to a significant extent by comparing hedonic house price predictions of each dwelling of a reference population for each period to the predictions for that house in the designated reference period \citep{waltl2019}.
For simplicity, we do not apply such an imputation approach, but use the fixed coefficient time dummy approach, similar to the benchmark case of \citet{waltl2019}.   We use the regressors: ``bedrooms'', ``bathrooms'', ``parking'', ``log land area'' and ``SA2 region''.  Unlike the granular D index version discussed below in Section~\ref{sec:compositional}, the log land area regressor is continuous and is not discretised.
An SA2 region is smaller than a SA3 region, so the SA2 region regressor used in this hedonic index is more granular than what we use in the D index. The other regressors bedrooms, bathrooms, parking and log land area allow this hedonic index to correct for compositional changes associated with these regressors. The D index in this paper does not use these variables as inputs, with the exception of the more granular index described below in Section~\ref{sec:compositional} where bedrooms and log land area in discretised form are used.
We estimate the conditional mean function as:

\begin{equation}
  \label{eq: linear_f}
\mathbb{E}(y_{it} \vert \mathbf{x}, \mathbf{D}) =  \mathbf{\beta}' \mathbf{x}_{it} + \sum_{s=0}^{T} \mathbf{D}_{ts} \delta_t 
\end{equation}

where $y_{it}$ denotes the sale price of dwelling $i$ at time $t$, $\mathbf{x}_{it}=(r_{it}, \mathbf{z}_{it})$ the vector of physical and locational characteristics of that dwelling. We note a slight abuse of notation since the the regional indicator $r_{it}$ is encoded as a dummy rather than entering the linear model directly. Also the vector of characteristics $]mathbf{z}$ includes a 1 to allow for an intercept. The matrix $\mathbf{D}_{ts}$ is a dummy variable that is $0$ except when $t=s$, taking the value of $1$. The $\mathbf{\beta}'$ and $\mathbf{\delta}$ vectors are the regression coefficients over the data set.

Generally, an hedonic index is constructed for times $1 \dots T$ using the $\hat{\mathbf{\delta}}=(\hat\delta_1,\dots,\hat\delta_T)'$ to represent additive growth in the mean of the log price relative to a reference time, say $t=1$, i.e. the growth in log price from time $1$ to $t$ is $\hat\delta_{t} - \hat\delta_{1}$. The (multiplicative) inflation in price is given by $e^{\hat\delta_{t} - \hat\delta_{1}}$. 
Dropping the normalisation at the reference time $t=1$, any constant $C$ in Equation~\ref{eq:hedonic_index} yields identical price inflation (and therefore an identical index after normalisation at $t=1$):
  
  \begin{equation}
  \label{eq:hedonic_index}
  H(t) = e^{\hat\delta_{t} + C }
  \end{equation}

Instead of only tracking price inflation, we want to examine the predicted average log price as well. Therefore, from Eq.~\ref{eq: linear_f}, the following choice for $C$ (and therefore dropping normalisation) tracks market averages:

\[
C=\sum_{t=1}^{T}\sum_{i=1}^{N_{t}} \mathbf{\beta}' \mathbf{x}_{it}
\]

\subsubsection{Repeat Sales Index}

The repeat sales method \cite{bailey:1963, k_case:1987}  avoids the need for detailed regressors by differencing out house features in an hedonic time dummy equation and regressing the observed price \emph{changes} per individual dwelling that sold at least twice, thus under-utilising the data. Although the index is intended to correct for compositional changes, time variations in house quality and shadow prices do affect it \citep{mcmillen2012}.  

Here, as a second benchmark, we also construct a Repeat Sales index. Naturally, we restrict the linear regression to the subset of sales data consisting of dwellings that have sold at least twice. Consider the log price change $y^{(t_2)}_i-y^{(t_1)}_i$ for repeat sale of house $i$ with log price $y^{(t_1)}_i$ in time period $t_1$ and log price $y^{(t_2)}_i$ in time period $t_2$ where $t_2>t_1$. Differencing Equation~\ref{eq: linear_f} with respect to consecutive dwelling sales and assuming no change in the characteristics encoded in $\mathbf{x}$, namely $\mathbf{x}_{it_2} - \mathbf{x}_{it_1} = \mathbf{0}$, we obtain:

\begin{equation}
  \label{eq: repeat_sales}
 \mathbb{E}(y^{(t_2)}_j-y^{(t_1)}_j) =  \delta_{t_2}-\delta_{t_1}
  \end{equation}

The matrix of regressors contains only time-related entries, with 1 in the column corresponding to the time period of the second sale and -1 and in the column corresponding to the time period of the first sale. 
Again, as can be seen from Equation~\ref{eq: repeat_sales} by exponentiation, the Repeat Sales index captures the appreciation of the geometric mean fitted output \citep[see also][]{shiller:1991}. 

The linear regression models for both the hedonic and repeat sales model contain a small ridge regularisation term. Also, unlike the D-indices, outliers have been removed from the training data using a random isolation forest algorithm. The linear indices are calculated for the 15th day of every month (e.g. 2022-02-15, 2022-03-15, ...): as such, the most recent date in our data is 2022-04-15.

\section{Results}
\label{sec:results}

The D-model generates a cross-sectional (log) price distribution for each spatial location at each time point. In this section we highlight the potential for the model to be used as a basis for exploratory data analysis and provide a number of sanity checks on the results. The model is trained on log prices and here the output functions have been transformed to functions defined on actual prices.

\subsection{The role of compositional changes}
\label{sec:compositional}

To examine the effect of compositional changes related to dwelling characteristics and a finer location resolution, we have also run experiments that resolve bedrooms, discretized land area and the SA2 subregions. We will refer to these experiments here as the ``granular'' experiments, and to the standard experiments, where SA3 subregion is resolved but not bedrooms and surface area, as the ``control'' experiments.

\begin{figure}[H]
\begin{center}
\scalebox{0.9}{\includegraphics[width=0.9\linewidth]{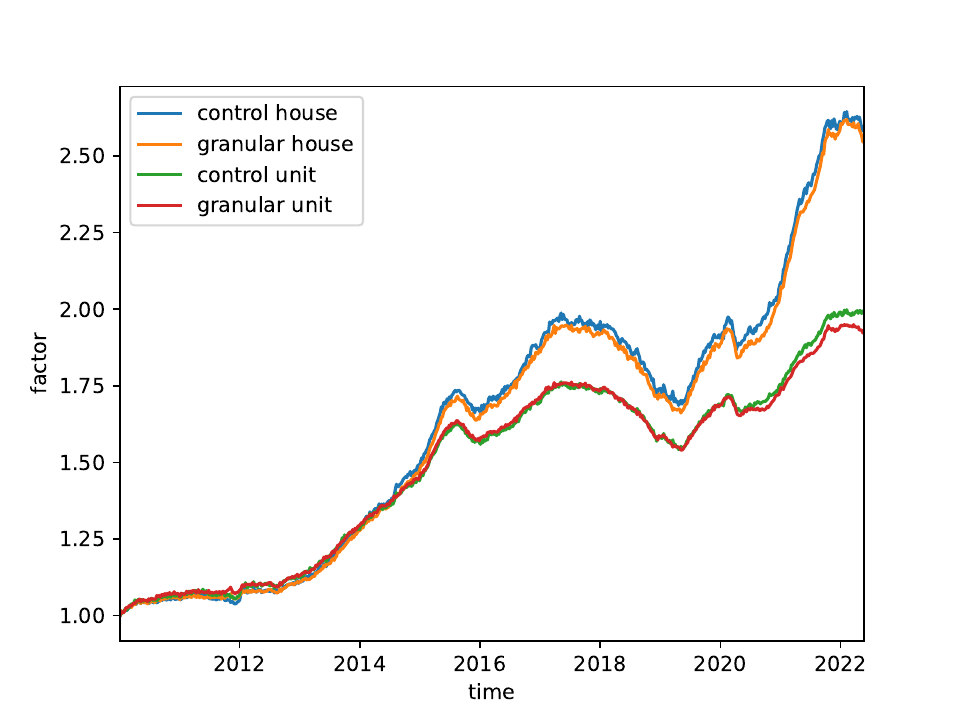}}
\end{center}
\caption{  The geometric mean price inflation since 1 Jan 2010 for Sydney over time for units in the control experiments (blue), units for the granular experiments (orange), houses for the control experiments (green) and houses for the granular experiments (red). The the granular index is on average 0.6\% below the control index for houses with a maximum departure of -3.3\%. For units this is 0.1\% resp. 3.6\% below the control index. }
\label{coarse_fine_graph}
\end{figure}

Figure~\ref{coarse_fine_graph} shows the geometric mean price inflation since 1 Jan 2010 for Sydney over time for apartment units and houses in the control and granular experiments. The experiments yield very similar results, with the exception of units for the most recent period, where the granular experiment yields lower values than the control experiment. This suggests that for units in Sydney, the most recent price increase has a component that is caused by compositional changes related to bedrooms (land area only varies for houses). This may relate to the recent apartment construction boom.

There is a possibility that the component Gaussian distributions of the mixture could be interpreted as subpopulations of unobservables. We have not examined this possibility in this study, leaving it for future study. However, we have conducted some experiments (not shown) where the component weights, or probabilities, of the Gaussian mixture $\alpha_k (\mathbf{x}, t;\mathbf{w})$ in Eq.~\ref{eq:mixture} are set constant to their average over the period Jan 1990 to Jan 2020. If the unobservable subcomponent interpretation holds, this fixing of the weights could control for compositional changes in these assumed unobservables. We found only small differences with the control D index experiment in Sydney. However, for units in Melbourne we found significant differences where the constant weighted experiment agrees significantly less well with a repeat sales index than the control experiment, leading to a less reliable index. This result suggests that fixing these weights does \emph{not} control for compositional changes in the assumed unobservables. However, this experiment remains somewhat inconclusive, as it does not account for the so-called ``labelling problem'', where permutations of the parameter set may yield the same mixture, e.g. when there are weights of similar values. Further development of this experimental procedure to manipulate the effect of assumed subpopulations of unobservables is left to future work and we will not consider fixing the mixture weights in this paper.

\subsection{Distributions}
\label{sec:result_pdf}

\begin{figure}[H]
\begin{center}
\scalebox{0.9}{\includegraphics[width=0.9\linewidth]{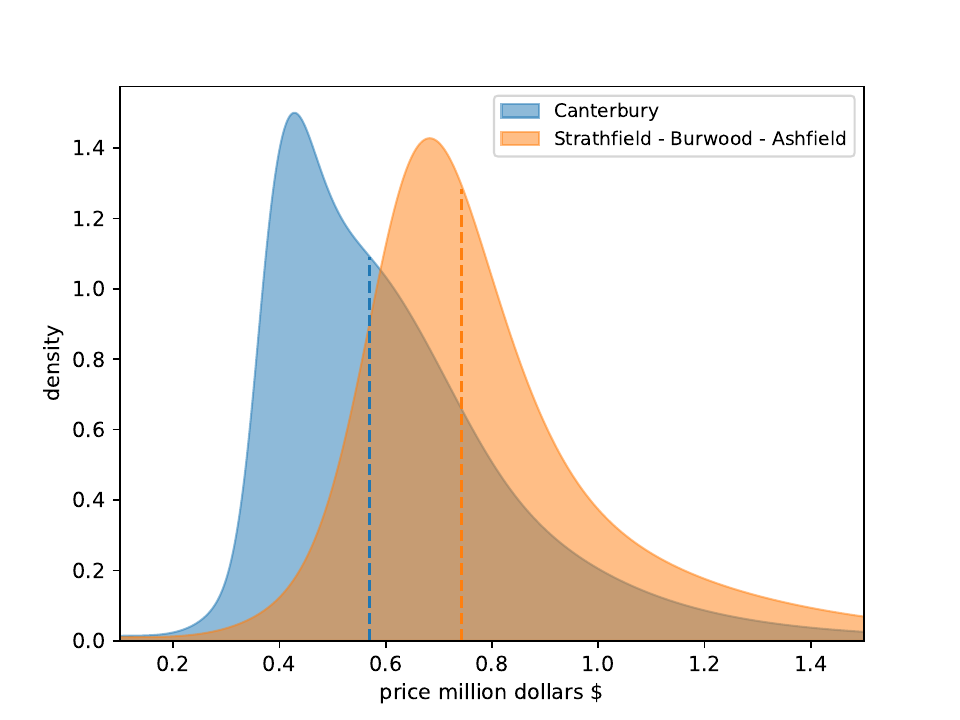}}
\end{center}
\caption{  The modelled PDFs for units as a function of sale price $y$ for the week of 2021-12-27 for two different (SA3) subregions: Strathfield - Burwood - Ashfield and Canterbury in Sydney.  The medians are displayed as vertical dashed lines with corresponding colors.}
\label{fig_burwood_canterbury_pdf_2022}
\end{figure}

To compare subregions, Figure~\ref{fig_burwood_canterbury_pdf_2022} shows the PDF for apartment units for the subregion `Strathfield - Burwood - Ashfield' and the subregion `Canterbury', both in Sydney for the week of December 27, 2022. These adjacent subregions are geographically close, yet exhibit distinct individual characteristics, such as a difference in median and overall shape of the PDF. Notable is the more near-bimodal structure of Canterbury: this could reflect geographical variation within this SA3 subregion.

\begin{figure}[H]
\begin{center}
\scalebox{0.9}{\includegraphics[width=0.9\linewidth]{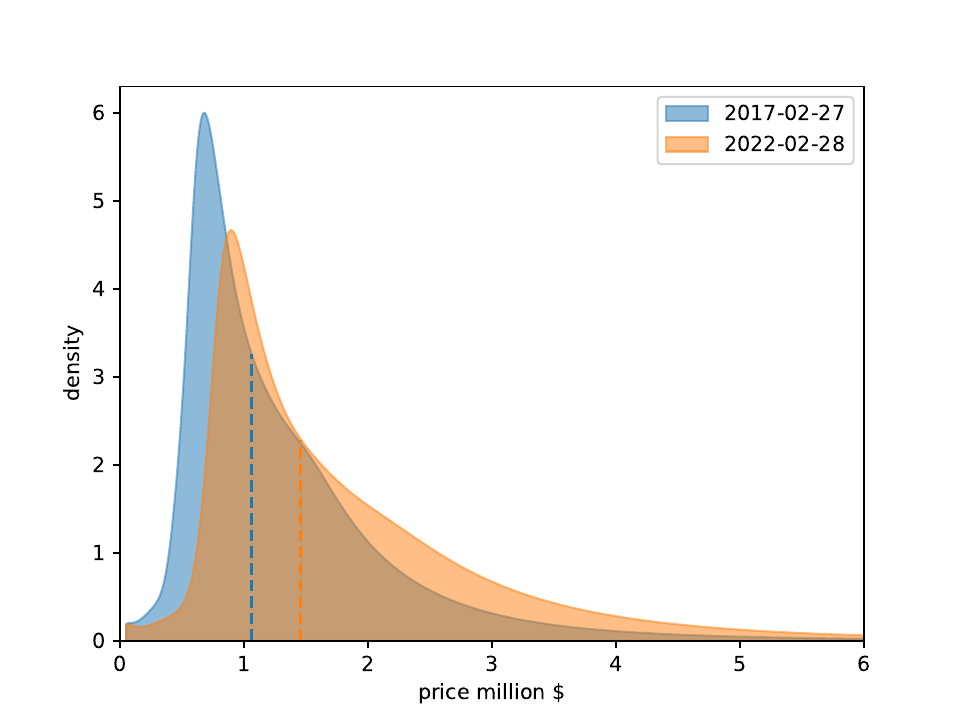}}
\end{center}
\caption{  The modelled PDF for Greater Sydney as a weighted mixture distribution of its individual component regional PDFs, shown for two points in time.  The medians are displayed as vertical dashed lines with corresponding colors.}
\label{pdf_sydney_composite}
\end{figure}

Figure~\ref{pdf_sydney_composite} shows the composite PDF for Greater Sydney as a weighted mixture of its individual component subregional PDFs at two different time points, February 27, 2017 and 28 February, 2022. It can be seen that in addition to the median price rising over the 5 year period, the right tail of the distribution of house prices has also become fatter. These insights could not be made by looking at hedonic and repeat sales indices. Also worth noting is that the composite PDF for all of Sydney is significantly more complex than the subregional component PDF (see Figure~\ref{fig_burwood_canterbury_pdf_2022} for comparison), and exhibits a broader price range. The median and mean major city indices below are taken from these major city level PDFs.

\subsection{Time series behaviour of several regions}
\label{sec:result_series}

\begin{figure}[H]
\begin{center}
\scalebox{0.9}{\includegraphics[width=0.9\linewidth]{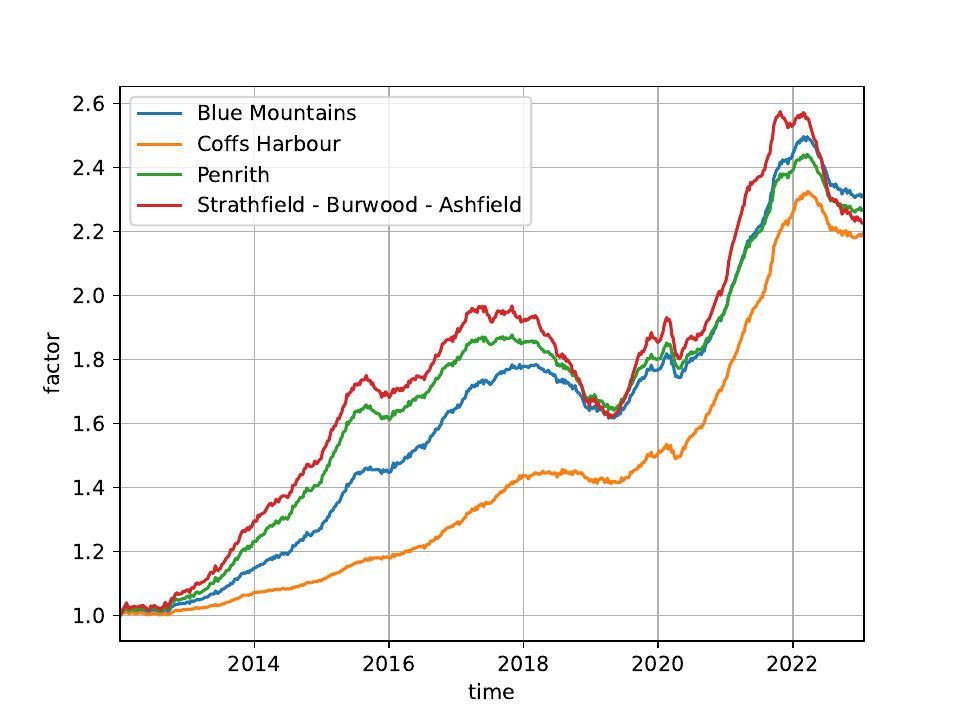}}
\end{center}
\caption{  Time series of median house price appreciation, indexed to 1 at 1 Jan 2012, derived from PDF time evolution for four different (SA3) subregions: Penrith, Strathfield - Burwood - Ashfield and Blue Mountains in the greater Sydney metro area and Coffs Harbour in coastal regional New South Wales. The vertical axis shows the factor by which to multiply the value at 1 Jan 2012 to obtain the index. }
\label{graph_bur_bm_penrith_coffs}
\end{figure}

To illustrate the price appreciation variations among regions captured by the indices based on our model, Figure~\ref{graph_bur_bm_penrith_coffs} shows time series of the D-median price index \emph{appreciation} since Jan 2012, derived from the PDF estimated by our model for the coastal town Coffs Harbour in regional NSW, and for several subregions within the greater Sydney metro area. This graph shows more recent data up to Jan 2013 to illustrate the interesting recent market behaviour. The Sydney market has undergone two periods of sharp growth, 2012-2017 and 2019-2022 and two periods of decline, 2018-2019 and 2022-present.

The model provides a detailed time-development for each region, with low levels of spurious noise. The model is able to capture in detail the substantially different time evolution of house prices between the regional coastal market of Coffs Harbour and the Sydney market. The internal differences in appreciation within the Sydney metro area further highlight the importance of the development of geographically granular indices. 

The Strathfield - Burwood - Ashfield area, located in Sydney's Inner West and closer to the inner city exhibits higher volatility than Penrith, a cheaper area located in the outer west (note that the graph shows growth, not actual price levels). In contrast to these two areas, the Blue Mountains population comprises significantly fewer commuters to the inner city due to longer travel times, causing a somewhat different appreciation behaviour. In particular, its appreciation lags behind up to around the beginning of 2017 and has caught up since then, exhibiting significantly more robust growth (compared to Sydney) from 2017 to present. This may be related to sustained increased demand due to the ``working from home'' trend much described in the media. The similarly attractive beach destination of Coffs Harbour, on the Mid North Coast of New South Wales and well outside Sydney's commuting catchment, exhibits this pattern more strongly, where growth lags considerably up to 2017 and then catches up, a trend continuing to the present.

Interesting granular behaviour is also observed, where the commencement of the sharp decline of 2022 is simultaneously timed for the four subregions around February/ March 2022, and a rate-of-decline reduction commencing around August 2022. The Strathfield - Burwood - Ashfield subregion is seeing the most significant decline and the present period of decline is also strongest for this subregion. Strictly speaking Strathfield - Burwood - Ashfield has its highest peak late 2021, whereas the other regions prolong subdued growth up to the sharp fall of Feb 2022. These patterns of recent growth and decline are also present at the metropolitan level index and are likely related to a prolonged period of very low interest rates followed by expectations around cash rate increases late 2021 (working through in fixed interest home loan rates) and the actual materialisation of significant rate increases by the Reserve Bank of Australia in 2022.

\begin{figure}[H]
\begin{center}
\scalebox{0.9}{\includegraphics[width=0.9\linewidth]{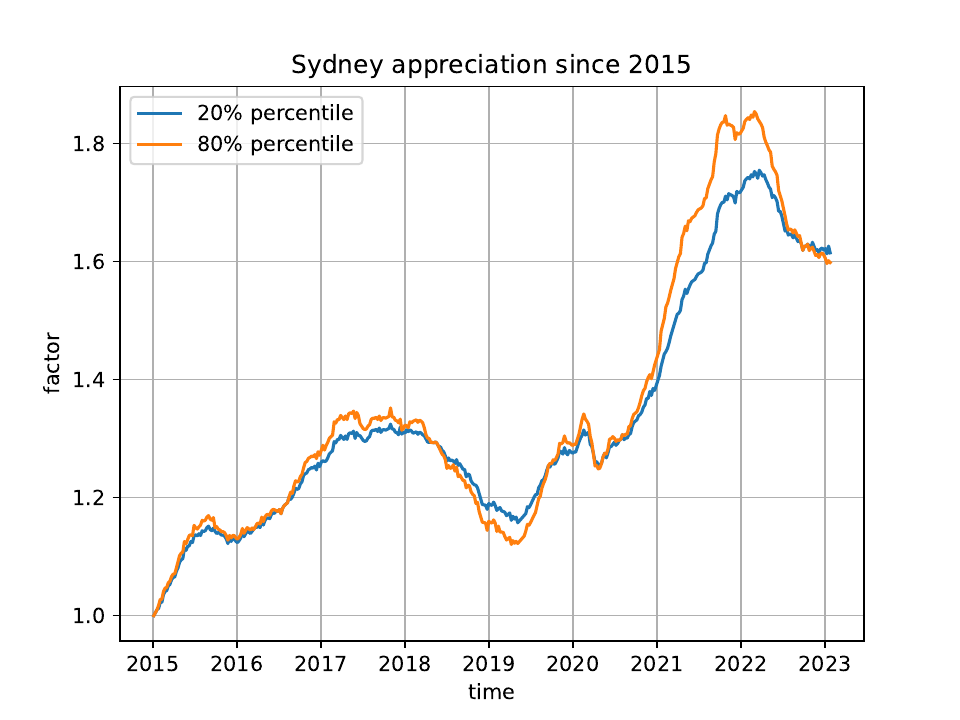}}
\end{center}
\caption{  Time series of price appreciation since Jan 2015 in Sydney for the 20\% and 80\% percentiles. As such, each index value is normalised to 1 for 1 Jan 2015. }
\label{quantile_growth_example}
\end{figure}

In addition to only the median, any quantile, or even multiple quantiles can also be used as an index. Similar to the quantile time series of \citet{waltl2019}, Figure~\ref{quantile_growth_example} shows the price \emph{appreciation} since Jan 2015 in Sydney for the 20\% and 80\% percentiles, with each index value normalised to 1 at 1 Jan 2015.  The more expensive 80 percentile group exhibits greater volatility over this recent period. In particular, the boom of 2020 and 2021 is significantly more pronounced for the expensive group, appreciating 90\%, compared to the cheap group, appreciating 80\%. The subsequent sharp decline is also more pronounced for the expensive group, for both the initial period of sharp decline in 2022 and the present (early 2023) period of milder decline. The effect of the commencement of the Covid19 pandemic in Australia is also clearly observed in early 2020 with a sharp high-frequency decline and subsequent recovery, taking more similar magnitudes for the expensive and cheap group. As the underlying metro-level PDF is built from the subregion PDFs, the quantile changes in the composite PDF can in principle arise from quantile changes in the component subregion PDFs as well as spatial variation in growth. 

\subsection{Time-persistence of individual dwelling price quantiles }
\label{cdf_test}

An interesting analysis that a distributional index allows for is to investigate the time persistence of price quantiles. For instance, a dwelling found in a low price quantile of its region is expected to retain a similarly low quantile value at a later sale. 
In contrast, random factors at the time of sale, subsequent renovations and changes in relative desirability over time between locations may introduce a regression to the mean where a dwelling is resold in a significantly different quantile (note that this is relative to a PDF that is changing in time), indicating that the desirability of the dwelling has changed compared to others.

Table~\ref{table:cdf_persistence} shows the correlation coefficients of the CDF (indicating the quantile) between two consecutive sales of the same dwelling relative to the metropolitan scale PDF (see for example Fig.~\ref{pdf_sydney_composite}) and the subregion scale PDF (see for example Fig.~\ref{fig_burwood_canterbury_pdf_2022}) for the three major cities, Sydney, Melbourne and Brisbane. 
A low correlation could indicate a market with highly changeable house characteristics and relative desirability or that the model is providing poor estimates of the true cross sectional price distribution.

The test shows a strong positive correlation of the quantiles between subsequent sales of the same dwelling. The Metropolitan scale quantiles have higher correlation coefficients. Indeed, the metropolitan PDF, comprising many subregions, has greater variance relative to dwelling-related noise. Correlation coefficients are expected to decrease with finer geographical scale as the PDF associated with the region becomes more specific and the signal to noise ratio relating to individual dwellings relative to the PDF decreases.  Also, as expected, the correlation is not perfect and there is a certain degree of regression to the mean. In conclusion, the model provides useful PDF estimates in the context of this test.

\begin{table}[h!]
\centering

\begin{tabular}{llrr}
\toprule
     City & Prop Type &  Metropolitan &  Subregion \\
\midrule
 Brisbane &     house &   0.90 &       0.79 \\
 Brisbane &      unit &   0.89 &       0.84 \\
Melbourne &     house &   0.89 &       0.70 \\
Melbourne &      unit &   0.88 &       0.78 \\
   Sydney &     house &   0.92 &       0.73 \\
   Sydney &      unit &   0.92 &       0.86 \\
\bottomrule
\end{tabular}

\caption{Correlation coefficients of the cumulative distribution function (CDF) between two consecutive sales of the same dwelling for the metropolitan scale PDF (column 3) and the subregion scale PDF (column 4). Shown are sales data between 2005 and 2022 for the three major cities, Sydney, Melbourne and Brisbane.}
\label{table:cdf_persistence}
\end{table}

\section{Model Validation}
\label{sec:performance}

We present alternative ways of validating the index relative to the benchmarks. These are motivated by the practical use of the model, namely the business task of estimating individual dwelling price inflation through price indexation, and whether the index reflects the true price quantiles in the market.

First, as a sanity check, we compare our proposed index to the hedonic index at the metro city scale, expecting them to be similar. Second, we will use index values to predict the growth of repeat sales dwellings. We use the D-index at a metropolitan scale and at a subregional scale comparing both to the linear indices. Our expectation is that predictions at the metropolitan scale will have a similar accuracy to those derived from benchmark indices, while those based on a subregional index should outperform the benchmarks. Third, we investigate whether the D-index is well calibrated in the sense that the empirical quantiles of sold homes closely match those of the index and tend to be conserved through time for individual dwellings. 
Note that, unlike above, the performance-related sections here use the data ending in May 2022.

\subsection{Comparison of the indices}

\begin{figure}[H]
	\begin{center}
		\scalebox{0.9}{\includegraphics[width=0.9\linewidth]{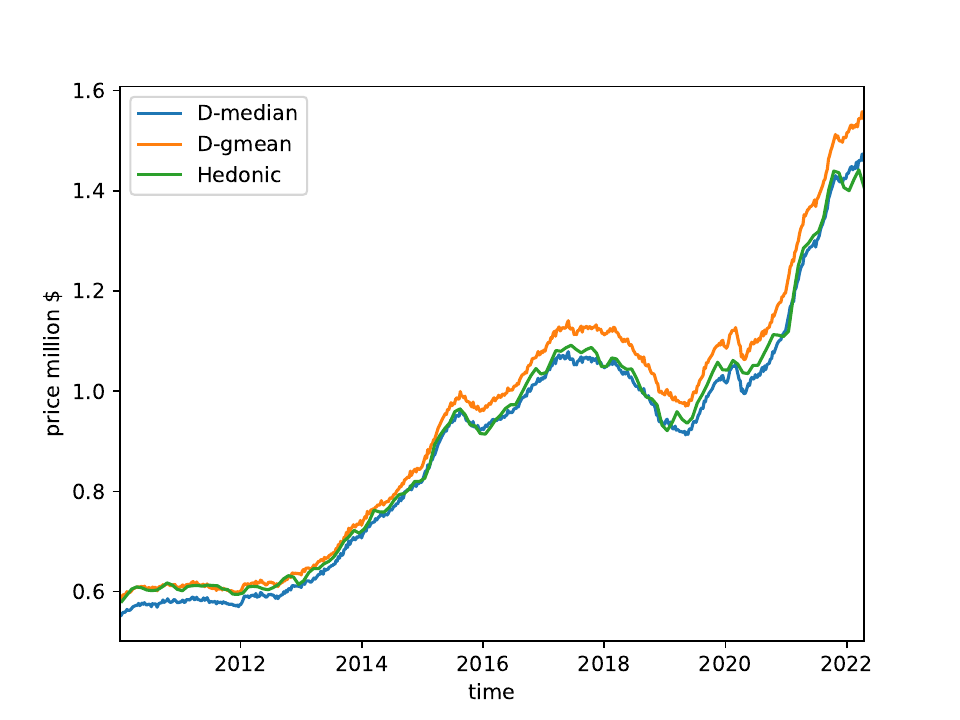}}
	\end{center}
	\caption{  Time series of the D-median index (blue), the D-gmean index (orange), and the Hedonic index (green) for houses in Greater Sydney. See Table~\ref{table:indices} for an explanation of the indices.}
	\label{graph_major_city_hedonic}
\end{figure}

Figure~\ref{graph_major_city_hedonic} shows the time series of the median house price D index and geometric mean price D index as a composite derived from the PDF evolution over time for underlying subregions for Greater Sydney. Despite being calculated using a different method, the indices are fairly similar to the hedonic index. The D-gmean and D-median indices are distinct from one another, allowing each index to be utilised in its own appropriate setting.  The Hedonic index displays greater volatility than the D-indices, including a mild seasonal signal associated with the Christmas period. In contrast, the D indices display very low spurious volatility, and do not exhibit a seasonal signal, such as a temporary drop in price over the Christmas period. 

\subsubsection{Comparison to distribution implied by Hedonic model}

\begin{figure}[H]
	\begin{center}
		\scalebox{0.9}{\includegraphics[width=0.9\linewidth]{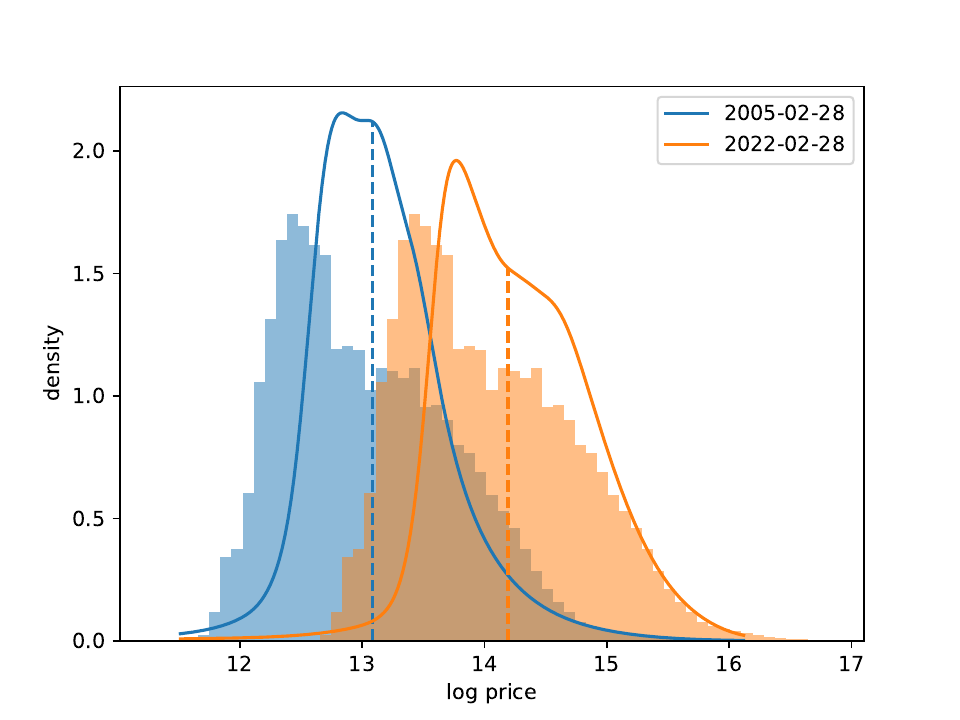}}
	\end{center}
	\caption{  Histograms of the collection of Hedonic model outputs $\hat{y}$, estimates of log price, over the entire training input set for two monthly time period with D model generated PDFs for the same periods overlaid for comparison. The vertical dashed lines indicate the medians of the D of the corresponding color.}
	\label{hist_hedonic}
\end{figure}

To further explore how our index is able to closely track the hedonic index, we compute fitted values of log price from the hedonic model for all houses sold in Sydney in two time periods, 2005-02-28 and 2022-02-28. These are plotted as histograms in Figure~\ref{hist_hedonic}. Overlaid on these histograms are the estimated distributions from our `D-model' for Sydney in the same time periods. It can be observed that the histograms and distributions share some similarities regarding their shapes, which suggests that by estimating the full distribution of log house prices, the D-model is able to `proxy' for the housing characteristics used in a hedonic model.

\subsection{Individual dwelling growth prediction accuracy }
\label{growth_accuracy}

 Here, we employ a k-fold validation test with 20 groups to examine the out of sample accuracy of the projected price for repeated sales to examine whether a subregion index is preferable in the practical business task of individual dwelling price indexation and to compare the D index to the linear indices. Namely, we conduct a repeat sales test to estimate the accuracy of each index as a tool to predict the price of the second sale of a home when the first sale is known. As such, we take dwellings that have sold at least twice, once at time $t_1$ for price $y^{(t_1)}$ and again at $t_2$ for $y^{(t_2)}$. Note that these are \emph{not} log prices, unlike Section~\ref{sec:dmodel}. 
  The projected price for repeat sale $j$ at $t_2$ is $ \hat{y}^{(t_2)} = \frac{H(t_2)}{H(t_1)} y^{(t_1)}_j$, where $H(.)$ is a time price index. We examine the accuracy of $ \hat{y}^{(t_2)} $ as an index-based estimate of the actual price $y^{(t_2)}$. 
  
  To obtain an out of sample error estimate, we use K-fold validation, where the dataset is divided into 20 groups and for each group the indices are fitted on the remaining 95\% of the dataset where that group is taken out, yielding 20 versions of the index. Sales in each group are predicted using the index corresponding to that group, so that each model has not been fitted on any of the validation data it is used on. As such, the validation data comprises the entire dataset of repeated sales.
  As indices we consider the D-gmean and D Subregion index and the hedonic and repeat sales index. 
  As error metrics we consider aggregates of the projection errors, namely the Median Absolute Percentage Error (MDAPE) is shown in column 3 under ``MDAPE projection'' and the Mean Absolute Percentage Error (MAPE) in column 4 under ``MAPE projection''. 

\begin{table}[h!]
\centering
\begin{tabular}{llrr}
\toprule
     City &        Index &  MDAPE &  MAPE \\
\midrule
 Brisbane &  D Subregion &   10.5 &  21.4 \\
 Brisbane &      D-gmean &   11.7 &  22.3 \\
 Brisbane &      Hedonic &   11.4 &  22.1 \\
 Brisbane & Repeat Sales &   11.9 &  23  \\
Melbourne &  D Subregion &   13.2 &  22.8 \\
Melbourne &      D-gmean &   14 &  23.7 \\
Melbourne &      Hedonic &   13.7 &  23.6 \\
Melbourne & Repeat Sales &   13.3 &  24.2 \\
   Sydney &  D Subregion &    9  &  15.9 \\
   Sydney &      D-gmean &    9.9 &  16.7 \\
   Sydney &      Hedonic &    9.8 &  16.6 \\
   Sydney & Repeat Sales &   10.4 &  17.6 \\
\bottomrule
\end{tabular}
\caption{Out of sample errors for individual dwelling growth predictions using the index between 2005 and 2020 for houses and units combined in greater metropolitan areas of Sydney, Melbourne and Brisbane.  Columns contain in successive order: index name, 
the Median Absolute Percentage Error (MDAPE) of the inflation error and the Mean Absolute Percentage Error (MAPE) of the inflation error. See Table~\ref{table:indices} for index explanation.}
\label{table:performance}
\end{table}

 We will focus on the MDAPE as it is less sensitive to outliers. The D-subregion index is more accurate than the metropolitan D-gmean indices in MDAPE. This shows that, when comparing the same methodology (namely the Gaussian mixtures), more localised subregion indices have lower median errors than their city-wide counterparts.
 The linear indices have similar MDAPE to the city-wide D-gmean index, with the exception of Melbourne, where the city-wide D-gmean has higher errors. This may explain why the D Subregion index is more on par with the linear indices there, as the Gaussian mixture methodology yields higher overall errors in that city. 
  The ``granular'' experiments of Section~\ref{sec:compositional} yield very similar performance metrics to the coarse experiments and are not shown here.

\begin{figure}[H]
	\begin{center}
		\scalebox{0.9}{\includegraphics[width=0.9\linewidth]{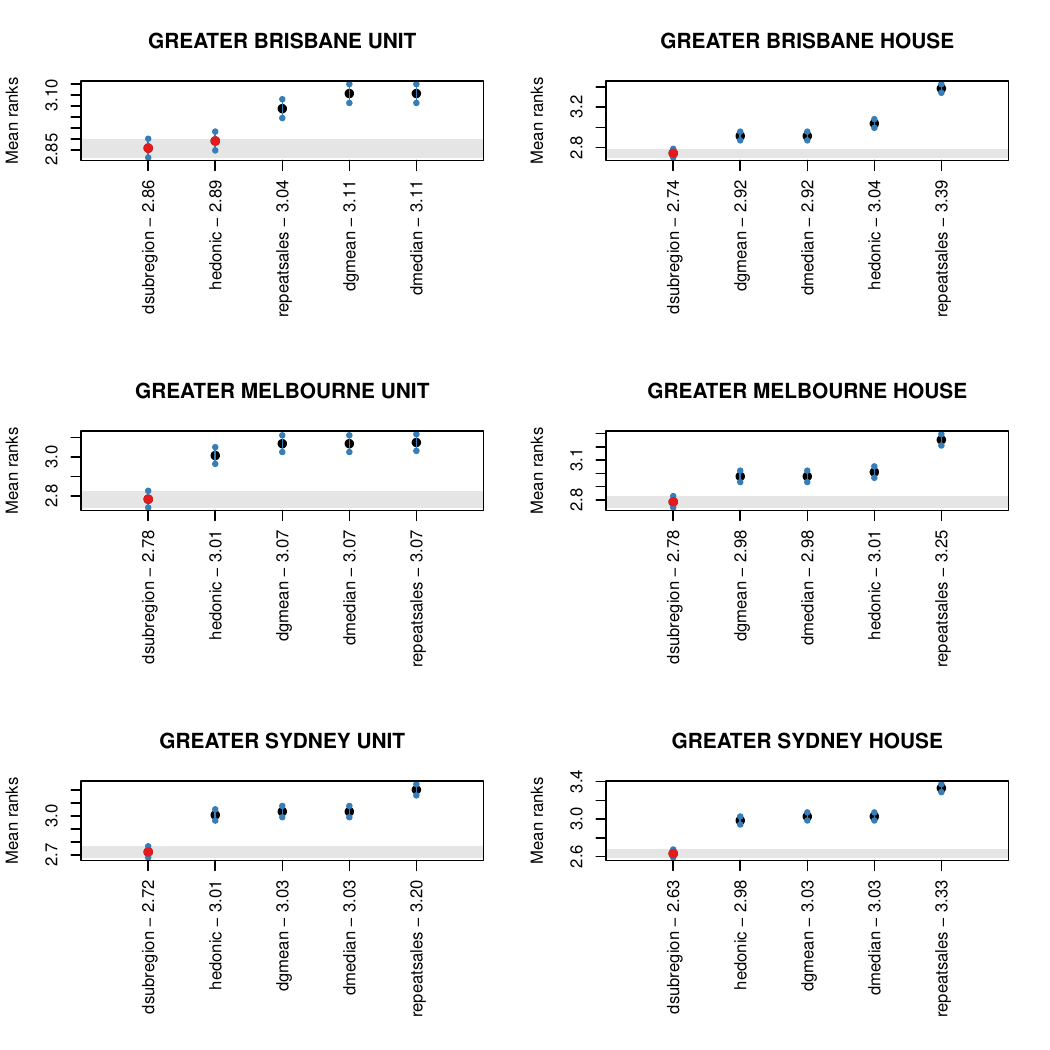}}
	\end{center}
	\caption{Visual presentation of post-hoc Nemenyi tests used to determine whether differences between prediction accuracy from using different indices is statistically significant.}
	\label{nemenyi_test}
\end{figure}

To determine whether the differences in prediction accuracy from Table~\ref{table:performance} are statistically significant, we consider non-parametric tests based on the ranks of absolute error. As part of the procedure, first a Friedman test is carried out to test the null hypothesis that there are no differences in the average ranking of absolute errors from using different indices. If the null is rejected (which occurs in all cases considered below), post-hoc Nemenyi tests are carried out to compare prediction accuracy from using different indices in a pairwise fashion, while controlling for multiple testing. The entire procedure is described in \cite{hollander:2013} and results are summarised as a Multiple Comparisons from the Best plot (see \cite{koning:2005}) in Figure~\ref{nemenyi_test}. In this plot, a method is statistically indistinguishable from the best method (in all cases ``D-subregion'') only if its confidence bands overlap with the confidence bands of the best method (shown in grey). This only occurs for the hedonic index and only in the case of units in Brisbane.

To examine whether the model without characteristics overfits on the data, we calculate the average negative log likelihood of the training data, $-0.218$ against out of sample data, $-0.219$: the similarity indicates that there is no overfit detected in this test.

\subsection{Calibration of the median and other quantiles}
\label{sec:median}

It is important to establish that the estimated distributions are well calibrated. This can be achieved by comparing the theoretical probabilities derived from the model to empirical probabilities. 

To examine how well an index reflects the median price, we calculate the percentage of sales at a price located below the index curve over the period under consideration. Over the long period under consideration where a substantial portion of the housing stock has sold at least once, the outcomes may be expected to be approximately equally distributed around an index that purports to reflect median price. As such, the D-median index is expected to be closest to the 50\% mark compared to the D-gmean and Hedonic indices, as these reflect geometric mean price and not the median. This is the pinball loss at the median and this will be generalised for other quantiles below in Equation~\ref{eq:pinball}. 

The result is shown in column $\delta$median in Table~\ref{table:delta_median}, indicating the absolute deviation from 50\% (e.g. if 51\% of the sales are below the curve, the deviation is 1\%). Unlike the other indices, the D-subregion index row represents the result for all the subregions (of the city), each according to their index, so here we report the median of these regional deviations from the median. The values have a 0.6 \% standard deviation with respect to experiments involving different load dates (see below).

\begin{table}[h!]
\centering
\begin{tabular}{llrrrr}
\toprule
     City &        Index &    $\delta$median \\
\midrule
 Brisbane &  D-subregion &                    1.2 \\
 Brisbane &      D-gmean &                      2.4 \\
 Brisbane &     D-median &                      0.3 \\
 Brisbane &      Hedonic &                      0.2 \\
 
Melbourne &  D-subregion &                    1.5  \\
Melbourne &      D-gmean &                      5  \\
Melbourne &     D-median &                    1.9 \\
Melbourne &      Hedonic &                    3.8 \\

   Sydney &  D-subregion &                    0.6 \\
   Sydney &      D-gmean &                     3  \\
   Sydney &     D-median &                    0.3 \\
   Sydney &      Hedonic &                     1.5 \\
  
\bottomrule
\end{tabular}
\caption{Estimates of how well each index approximates the median price. Columns contain in successive order: index name and deviation from 50\% in percentage of sales below the index curve index. See Table~\ref{table:indices} for index explanation. Data is between 2005 and 2020 for houses and units combined in greater metropolitan areas of Sydney, Melbourne and Brisbane. See Table~\ref{table:indices} for index explanation. }
\label{table:delta_median}
\end{table}

The D-median index is closest to 50\% among the metropolitan indices in Sydney and Melbourne.  This lends support to the D-median index being more representative of a median than the Hedonic and D-gmean indices. There is no significant distinction between the indices in Greater Brisbane based on this metric. This is because in Brisbane, the metropolitan indices follow values that are more similar to each other (figure not shown).
Separately, the D-subregion index also reflects the median reasonably well on this metric at the regional level.

\begin{figure}[H]
\begin{center}
\scalebox{0.9}{\includegraphics[width=0.9\linewidth]{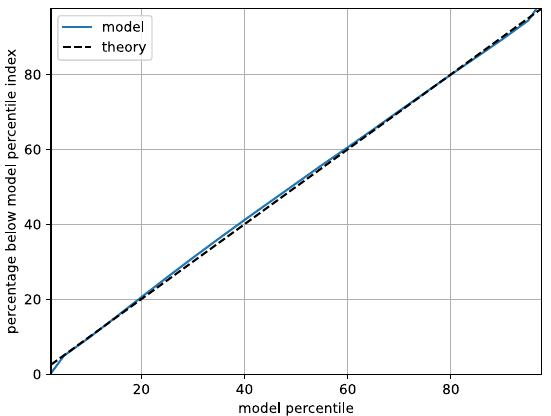}}
\end{center}
\caption{  Percentage of dwelling sales below the corresponding percentile curve predicted by the Greater Sydney composite D model. The horizontal axis contains the percentile tested for and the vertical axis shows the percentage of sales data found below that percentile as predicted by the model. See Table~\ref{table:delta_median} for the 50\% (median) percentile, corresponding to the D-median index. The data comprises houses and units and the period is between 2005-01-01 and 2022-01-01. }
\label{percentiles_test}
\end{figure}

The D model provides quantile estimates (see Fig~\ref{quantile_growth_example}). Therefore, the above test for the median can be generalised to any percentile $\pi$. Consider the percentage $\hat{\pi}$ of observed sales that lie below the $\pi$-percentile curve $y_{\pi, t,r}$ as derived from the D model, where $r$ signifies a specific subregion that encloses the location of sale $i$. Here we consider only the metro scale PDF so that the region subscript $r$ indicates a single metropolitan area (namely Sydney) and the subscript could be dropped in this case. We can compute this:

\begin{equation}\label{eq:pinball}
\hat{\pi}=\frac{1}{N}\sum\limits_{t=1}^T\sum\limits_{r=1}^R\sum\limits_{i\in\mathcal{I}_{t,r}}^R I(y_i<y_{\pi, t,r})
\end{equation}

where $I(.)$ is a function equal to one if the statement inside the parentheses is true and zero otherwise. Equation~\ref{eq:pinball} is related to the pinball loss function, and for a well calibrated model, the observed $\hat{\pi}$ should be close to the applied $\pi$. Figure~\ref{percentiles_test} plots $\hat{\pi}$ against $\pi$  for different values of $\pi$, for Sydney.  For a well calibrated model, the line should be close to a 45 degree line. The fact that these lines are almost indistinguishable is a testament to the accuracy of the D-model.

\subsection{Effect of data sparsity}
\label{sparsity}

\begin{figure}[H]
\begin{center}
\scalebox{0.9}{\includegraphics[width=0.9\linewidth]{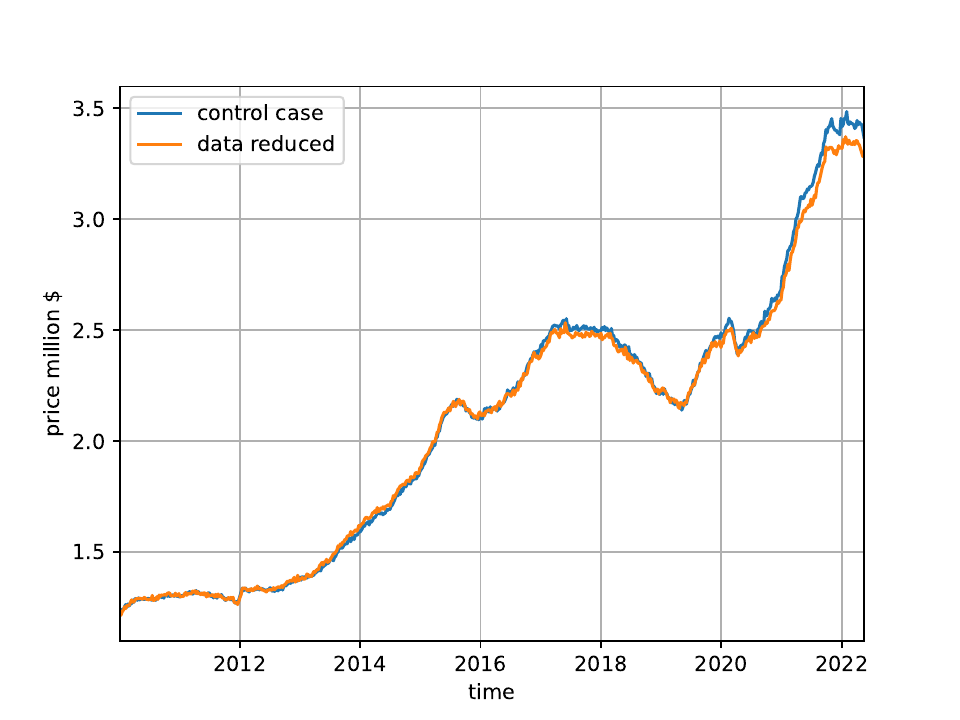}}
\end{center}
\caption{   Geometric mean indices for for subregion Chatswood - Lane Cove from Jan 2010 for the unmodified model, the control case in blue, and the modified model where 90\% of the data for the subregion is removed, orange. The difference in 2022 is around 3\%.}
\label{fig:sparse_neighbor}
\end{figure}

The model has been constructed so as to predict for a subregion along with its neighbours. As a result, we expect the model to show relatively robust results for a subregion when the data volume for that particular subregion is deliberately reduced. Here, we show the results of this experiment for the randomly selected subregion Chatswood - Lane Cove. This experiment requires a ``control'' case, the unmodified index for this subregion, and a ``treatment'' case, where we have fitted the model, again in ensemble, on data where we retain only 10\% of that data for this subregion (and so removed 90\%). Figure~\ref{fig:sparse_neighbor} shows that the results of the data-sparse treatment case are similar to the control case, with the greatest departure in the market-boom of the most recent years. This difference is non-negligible yet not very large.

\section{Discussion}
\label{sec:discussion}

In this paper we introduce a model estimating the parameters of a time-evolving Gaussian mixture to approximate the conditional cross-sectional house price distribution of each unique combination of house characteristics and locational features, namely property type and subregion. Hierarchically, price distributions of agglomerations of these feature combinations, e.g. metropolitan areas as made up of subregions, are constructed as a weighted sum of component distributions where weights are time-invariant.
This study adds the use of Gaussian mixtures to the emerging body of work that aims to compute the cross-sectional house price density distribution to gain a more complete view of the housing market. 
The fitting procedure is carried out by a deep network. We expect that alternative fitting procedures could also work. Although not essential, beneficial in this context would be a model where neighbours of regions are taken into account. 

We have demonstrated that the ensemble application of our model yields low-noise stable price distributions and indices defined on the granular submetropolitan scale of the Australian SA3 regional level. 
Comparing the D-index at the city level to the subregion level in the three main cities, we find that these more geographically localised indices yield lower prediction errors in the practical task of individual dwelling price projection based on previous sales values. This lends further support to previous findings that a city-wide index is not sufficient to capture the various growth characteristics covered by the average price. 
The modelled distribution quantiles agree well with the the data and we we confirm previous findings that different quantiles evolve differently over time.

Using this deep network model, we construct stable low-noise indices using relatively few house characteristics and locational features, only property type and location, demonstrating its potential usefulness in data-poor environments with few features available. The Gaussian mixture allows a range probability densities, thus reducing the bias associated with unresolved house characteristics (we do not examine these subpopulations in this paper). If required and available, features can be added to construct finer granular density functions, providing more detailed insights and a more explicit correction for compositional changes.

\newpage


\clearpage
\begingroup
\bibliographystyle{elsarticle-harv}
\setlength\bibsep{0.5pt}
\bibliography{new_ref}
\endgroup

\appendix
\clearpage

\end{document}